\documentclass[preprint2]{aastex}

\def\vlsr{$V_{\mbox{\scriptsize LSR}}$}
\def\kms{~km~s$^{-1}$}

\def\etal{~et\ al.\ }
\def\h2o{H$_{2}$O}
\def\iras1913{IRAS~19134$+$2131}
\def\masyr{~mas~yr$^{-1}$}
\def\j1925{J1925$+$2106}

\slugcomment{To be submitted to the Astrophysical Journal.}

\shorttitle{3D kinematics and distance of \iras1913.}
\shortauthors{H.~Imai, R.~Sahai, \&  M.~Morris}

\begin{document}

\title{The spatio-kinematical structure and distance of the pre-planetary 
nebula \iras1913}

\author{Hiroshi Imai}
\affil{Department of Physics, Faculty of Science, Kagoshima University, 
1-21-35 Korimoto, Kagoshima 890-0065, Japan}
\email{hiroimai@sci.kagoshima-u.ac.jp}

\author{Raghvendra Sahai}
\affil{Jet Propulsion Laboratory, 4800 Oak Grove Drive, Pasadena, CA 91109}
\email{raghvendra.sahai@jpl.nasa.gov}

\and 

\author{Mark Morris}
\affil{Department of Physics and Astronomy, University of California, 
Los Angeles, CA 90095$-$1547}
\email{morris@astro.ucla.edu}

\received{2007 April 10}

\begin{abstract}
Using the {\it Very Long Baseline Array} at six epochs, we have observed \h2o\ maser emission in the pre-planetary nebula \iras1913 (I19134), in which the \h2o maser spectrum has two groups of emission features separated in radial velocity by $\sim$100\kms. We also obtained optical images of I19134 with the {\it Hubble Space Telescope} to locate the bipolar reflection nebula in this source for the first time. The spatio-kinematical structure of the \h2o\ masers indicates the existence of a fast, collimated (precessing) flow having a projected extent of $\sim$140~mas and an expansion rate of $\sim$1.9\masyr\ on the sky plane, which gives a dynamical age of only $\sim$40~yr. The two detected optical lobes are also separated by $ \sim$150~mas in almost the same direction as that of the collimated flow. The good agreement between the extent and orientation of the \h2o\ maser outflow and optical lobes suggests that the lobes have been recently formed along the collimated fast flow.  Thus the circumstellar envelope around the evolved star has apparently been penetrated by the fast flow and has been cleared to form the way for the emergence of the starlight in the directions of the fast flow. The positions of all of the detected maser features have been measured with respect to the extragalactic reference source \j1925 over one year. Therefore we analyzed maser feature motions that consist of the combination of an annual parallax, a secular motion following Galactic rotation, and the intrinsic motions within the flow. We obtain an annual-parallax distance to I19134 of  $D=8.0^{+0.9}_{-0.7}$~kpc and estimate its location in the Galaxy to be $(R, \theta, z)=$($7.4^{+0.4}_{-0.3}$~kpc, 62\arcdeg$\pm 5\arcdeg$, $0.65^{+0.07}_{-0.06}$~kpc). From the mean motion of the blue-shifted and red-shifted clusters of maser features, we estimate the 3-D secular motion of I19134 to be $(V_{R}, V_{\theta}, V_{z})=
(3^{+53}_{-46}, 125^{+20}_{-28}, 8^{+48}_{-39})$ [km~s$^{-1}$]. From the height from the Galactic plane, $z$, and the velocity component perpendicular to the Galactic plane, $V_{z}$, we estimate a rough upper limit of $ \sim$9~$M_{\sun}$ 
to the stellar mass of I19134's progenitor. 
\end{abstract}


\section{Introduction}
A stellar jet appearing at the final stage of stellar evolution is a manifestation of the important, but still poorly understood phenomenon leading to the formation of asymmetric planetary nebulae (PNe). Bipolar and multipolar optical morphologies and the common presence of point symmetry are frequently found in a sample of young PNe. Based on this fact, \citet{sah98} concluded that collimated jets have been operational in a large fraction of PNe, and are one of the major factors shaping these objects. Similar morphologies have been found in a survey of pre-planetary nebulae (PPNe, \citealt{sah04}; Sahai \etal\ 2007 in preparation), leading to the current paradigm that such a collimated jet is launched near or at an earlier phase of stellar evolution  -- the asymptotic giant branch (AGB) phase - and is frequently observable in molecular line emission. 

There are a small number of candidate sources that show us the earliest stage of stellar jet emergence. ``Water fountain sources`` are the most promising candidates; they are stellar objects that have extremely high-velocity flows traced by \h2o\ maser emission. The outflow velocity sometimes exceeds 100\kms, much greater than the typical expansion velocity of circumstellar envelopes (CSEs) of Mira variable and OH/IR stars traced by 1612-MHz OH maser emission (10--25\kms, e.g., \citealt{lin89}). About 10 water fountains have been identified in previous single-dish observations so far: 
IRAS~15445$-$5449 $=$ OH~326.5$-$0.4; IRAS~15544$-$5332 $=$ OH~325.8$-$0.3; IRAS~16342$-$3814 $=$ OH~344.1$+$5.8; IRAS~16552$-$3050 $=$ GLMP~498; IRAS~18043$-$2116 $=$ OH~0.9$-$0.4; IRAS~18139$-$1816 $=$ OH~12.8$-$0.9; IRAS~18286$-$0959; IRAS~18450$-$0148 $=$ W43A, OH~31.0$+$0.0; IRAS~18460$-$0151 $=$ OH~31.0$-$0.2; IRAS~18596$+$0315 $=$ OH~37.1$-$0.8; \iras1913, hereafter abbreviated as I19134  (e.g., \citealt{gen77,bau79,eng86,lik92, eng02, dea07, deg07}, and references therein). The spatio-kinematical structures of most of them have been revealed by the high angular resolution of radio interferometers (e.g., \citealt {dia85,gom94,sah99,ima02,mor03,ima04}, hereafter Paper {\rm I}; 
\citealt {cla04,ima05}, hereafter INDMD05; \citealt{bob05,vle06}). VLBI observations of some of these \h2o\ maser sources have revealed that the jets are highly collimated and have extremely short dynamical ages ($\lesssim$100~yr, \citealt{mor03}: Paper {\rm I}; INDMD05; \citealt{bob05}). 

Collimated bipolar outflows in {\it young} stellar objects are perhaps in atomic or ionized form as a result of the violent acceleration processes which brought the outflows to very high velocities. On the other hand, the molecules are believed to form in the compressed, rapidly cooling, post-shock regions at the interface between the jets and ambient molecular clouds (e.g, \citealt{bac96}). A similar interaction is expected between a fast stellar jet and the ambient, slow AGB outflow (e.g., \citealt{lik92, san01}). In either case, molecular emission in stellar jets, as seen in \h2o\ maser emission, is strongly collimated both geometrically and kinematically. This implies that high density gas, which may not be originally molecular, is supplied from the very vicinity of the stellar surface or the innermost part of the CSE to the jet tips. In the water fountain source W43A, precession of the stellar jet is also confirmed (INDMD05). \citet{vle06} found that the \h2o\ maser emission in W43A exhibits strong Zeeman splitting and linear polarization, strongly suggesting that the W43A jet is driven magneto-hydrodynamically. The launching of such jets apparently takes place at the end of the AGB phase or the beginning of the post-AGB phase during which stellar mass loss is still most active. 

The \h2o\ maser emission in the ``water fountains'' is located over 100~AU away from the central stellar objects, so it does not shed any light on regions close to the stars. It is still unclear whether the dynamical center of the jet coincides with the evolved star itself. However, the symmetrical placement about the central star as seen in many planetary and pre-PNe suggests that such jet-induced shock structures -- which are likely to be related to the \h2o maser spot clusters in the water fountain sources -- are both spatially and dynamically centered on the star.  Note that, on the other hand, a molecular stellar jet traced by \h2o\ maser emission will disappear within 1000~yr or shorter because of the imminent exposure of the PN nucleus that causes photo-dissociation of \h2o\ (c.f., \citealt{mir01}). The number of water fountain candidates is increasing, so the duration of the water fountain phase should be better determined statistically and kinematically. Within this short period, the spatio-kinematical structures of the \h2o\ maser emission should undergo substantial evolution, which might be directly confirmed by long-term observations within a human lifetime. 

In this paper, we present the spatio-kinematics of \h2o\ masers associated with I19134, as revealed by six-epoch observations with the {\it Very Long Baseline Array} (VLBA), together with the optical morphology obtained by the {\it Hubble Space Telescope} (HST). The preliminary result obtained with the VLBA data in the first two epochs was published in Paper {\rm I}, but the data from all epochs were reduced in the same procedure, although with additional data calibration in the present paper. With the new data, we better determined the dynamical age of the I19134 flow and the location of I19134 in the Galaxy using the phase-referencing technique in the VLBA observations. \S 2 describes in detail HST and VLBA observations and data reduction. \S 3 describes 
the results. \S 4 discusses the implications of the maser spatio-kinematics for the evolutionary status of I19134 and of the astrometry to determine the distance to I19134 and its location and origin in the Galaxy. In the present paper, a distance to I19134 of 8.0~kpc (see section \ref{sec:distance}) is derived and adopted to infer other physical parameters of I19134. 

\section{Observations and data reduction}

\subsection{HST observation}
\label{sec:HST}
I19134 was imaged on UT date 2002 December 19 (GO program 9463), by the Wide Field Camera (WFC) of the Advanced Camera for Surveys (ACS), which has a plate scale of 0\farcs05~pixel$^{-1}$. Two dithered exposures of 338~s each were obtained through each of the two filters (F606W, pivot wavelength$=$0.599~\micron, and F814W, pivot wavelength=0.806~\micron). The standard STScI/HST pipeline calibration has been applied to all data. We have preserved the intrinsic orientation of the HRC images in the figures in this paper, electing not to rotate them for alignment of their horizontal and vertical axes with the cardinal directions. This is because such rotation would result in some degradation of the image quality. Several field stars present in the images have been used to achieve satisfactory registration between the dithered images, which were sub-sampled by a factor of two before being registered and combined to produce the final images. Cosmic-ray hits were also removed during this process. The angular resolution in the F606W image, as measured from many field stars, is 0\arcsec.11.

\subsection{VLBA observations of the \h2o\ masers}
\label{sec:VLBA}

\begin{table*}[ht]
\caption{Parameters of the VLBA observations and data reduction for individual epochs.}\label{tab:status}

\scriptsize
\begin{tabular}{lccrcrcc} \hline \hline
Observation & Epoch & Duration  
& Recording & rms noise & \multicolumn{1}{c}{Beam\tablenotemark{2}}
& Detected & Remark\tablenotemark{3} \\
code & (yy/mm/dd) & (hr) & mode\tablenotemark{1} & (mJy beam$^{-1}$) 
&  \multicolumn{1}{c}{(mas)} & features & \\ \hline 
BI25A \dotfill & 03/01/04 & 8 & 4$\times$4~MHz & 8.6 
& 1.07$\times$0.40, $-$9.4$^{\circ}$ & 21 & $\circ$ \\
BI25B \dotfill & 03/03/07 & 8 & 4$\times$4~MHz & 8.6  
& 0.92$\times$0.39, $-$13.4$^{\circ}$ & 28 & $\circ$ \\
BI28A \dotfill & 03/10/25 & 6 & 2$\times$8~MHz & 11.2  
& 0.84$\times$0.37, $-$5.3$^{\circ}$ & 16 & $\circ$ \\
BI28B \dotfill & 03/12/20 & 6 & 2$\times$8~MHz & 11.6  
& 0.97$\times$0.35, $-$11.0$^{\circ}$ & 15 & $\times$ \\
BI28C \dotfill & 04/02/09 & 6 & 2$\times$8~MHz & 11.3  
& 0.81$\times$0.36, $-$3.0$^{\circ}$ & 22 & $\circ$ \\ 
BI28D \dotfill & 04/04/26 & 6 & 2$\times$8~MHz & 8.5  
& 0.81$\times$0.36, $-$5.1$^{\circ}$ & 12 & $\circ$ \\ \hline
\end{tabular}

\tablenotetext{1}{Number of base-band channels and bandwidth per BBC}
\tablenotetext{2}{Synthesized beam size resulting from natural weighted visibilities; 
major and minor axis lengths and position angle.}
\tablenotetext{3}{Correction for atmospheric zenith delays: valid 
($\circ$) or invalid ($\times$).}
\end{table*}

Table \ref{tab:status} summarizes the status of the VLBA observations of the I19134 \h2o\ masers and data reduction. The VLBA observations were made at six epochs spanning the period from 2003 January 4 to 2004 April 26 (BI~25 and 28). The duration of each observation was 8--10~hr in total, including scans on the calibrators NRAO~512 and J2148$+$0657, which were observed for 6 minutes every 45 minutes for calibration of the bandpass characteristics. The phase-referencing mode was adopted, in which each antenna nodded between the phase-reference and target maser sources in a cycle of 60~s. The on-source duration in each of the cycles was shorter than 20~s, depending on the position angle of the target--reference separation and on antenna zenith angle. In the first two epochs, the sources J1910$+$2305 and J1925$+$2106 (hereafter abbreviated as J1925) were observed as phase-references, while at other epochs only the latter was observed. In this paper, we show the astrometric results obtained with the data of J1925, which is $\sim$2\arcdeg.5 away from I19134 and was well detected in every antenna-nodding cycle. As a result, the effective coherent integration time of the maser data was $\sim$2.5~hr at all epochs. The received signals were recorded at a rate of 128~Mbits~s$^{-1}$ with 2 bits per sample into four or eight base-band channels (BBCs) in dual circular polarization. Two of the BBCs covered each of the velocities of the red-shifted and blue-shifted maser components of I19134, respectively. The recorded data were correlated with the Socorro FX correlator using an integration period of 2~s. The data of each of the BBCs were divided into 256 or 512 spectral channels, corresponding to a velocity spacing of 0.2\kms\ per spectral channel. The following coordinates of I19134 were adopted as the delay-tracking center in the data correlation:
$\alpha_{J2000}=$19$^{h}$15$^{m}$35$^{s}$\hspace{-2pt}.2150,
$\delta_{J2000}=$~$+$21$^{\circ}$36$^{\prime}$33\farcs 900.

For the VLBA data reduction, we used the NRAO's AIPS package and applied the procedures for the phase-referencing technique (e.g., \citealt{bea95}). Note that, at the last four epochs, the data correlation was made with the Earth orientation parameters (EOPs), the Earth's pole position and UT1$-$UTC, predicted about two weeks before the observation date. This makes a difference of tens of milliarcseconds in the observation geometry from that estimated from the EOPs obtained in the actual geodetic VLBI observations \citep{wal05}. The error in the target position {\it with respect to the reference source} may be smaller than the value mentioned above, or the error in the {\it absolute coordinates} of the reference source, but may not be negligible.
First, therefore, the visibilities obtained from the original correlation were corrected as if the delay-tracking is made using the EOPs precisely estimated. Second, residual delay/delay-rate solutions were obtained from fringe fitting for scans on the calibrators. In this stage, we estimated that the residuals in the delay solutions had uncertainties less than 3~nsec. Then fringe fitting was carried out for J1925 scans. Most of the residual delay-rate solutions were smaller than 10~mHz; we could thus avoid any 2$\pi n$ radian ambiguity in the fringe-phase interpolation between successive scans on J1925. All of the calibration solutions obtained for J1925, including the self-calibration solutions, were applied to the maser data. We used a synthesized beam based on naturally weighted visibilities to make image cubes. The beam parameters and the rms noise level in the maser cubes (in spectral channels without bright maser emission) are summarized in table \ref{tab:status}. 

We determined the position of a maser feature (a physical maser clump consisting of a cluster of maser spots or velocity components located at almost the same position and velocity) by calculating an intensity-weighted mean of positions of maser spots in the feature. Table \ref{tab:features} lists all maser features detected in the observations. The feature positions are measured with respect to the delay-tracking center whose coordinates are given above. The maser astrometry may be affected by the brightness structure of J1925, but the effect of that is estimated to be less than 15~$\mu$as (Paper {\rm I}).  Although I19134 is so distant that apparent angular structures of the individual maser features are relatively compact and simple, the low signal-to-noise ratios could still affect the feature position measurement at a level less than 70~$\mu$as. A relative difference in the excess path delays in the atmosphere between the reference and the maser sources also occurs due to different zenith angles at each station. Errors in position measurements due to unknown residual atmospheric zenith delay is described in Paper {\rm I} and references therein. In the present paper, we attempted correction for the residual zenith delay as follows. First, we obtained a maser map using the normal procedures for the phase-referencing technique mentioned above. The maser map obtained was used as a trial map to subtract phase variation with time from the visibility phases used for the maser mapping. The residual visibility phases are fit to modeled curves expected from given residual zenith delays with their linear drifts with time. In the model fitting, uncertainties in the estimated residual zenith delays were typically 5~ps (1.5~mm in length). For the maser--reference pair separated by 2\arcdeg.5 in the present paper, this uncertainty translates to a difference of $ \sim$0.4~ps ($ \sim$120~$\mu$m in length) in the excess path delays towards the individual sources at a mild zenith angle ($z\sim$50\arcdeg).  This, in turn, corresponds to a position error of 26~$\mu$as when observing with a 1000-km baseline (a fringe spacing of 2.78~mas at $\lambda=$1.35~cm). The correlated intensity in the maser images after this correction was also increased by 10--15\% from the trial maser images. In conclusion, the positional accuracy of a maser feature is typically about 50~$\mu$as. For BI28B, however, the accuracy was worse because the solutions for the excess path delay correction were invalid due to poor weather conditions that made the model fitting difficult. 

\begin{table*}[ht]
\tablenum{3}
\caption{Relative and absolute proper motions of 
\h2o\ maser features.}\label{tab:pmotions}

\scriptsize
\begin{tabular}{lrrr@{ }rr@{ }rr@{ }rr@{ }rr@{ }r} 
\hline \hline 
Maser & \multicolumn{2}{c}{Position}
 & \multicolumn{4}{c}{Relative}
 & \multicolumn{4}{c}{Absolute}
 & \multicolumn{2}{c}{Radial} \\  
feature\#\tablenotemark{1} & \multicolumn{2}{c}{Offset\tablenotemark{2}}                         
 & \multicolumn{4}{c}{Proper motion\tablenotemark{3}}
 & \multicolumn{4}{c}{Proper motion}
 & \multicolumn{2}{c}{motion\tablenotemark{4}} \\    
 & \multicolumn{2}{c}{(mas)}
 & \multicolumn{4}{c}{(\masyr)}
 & \multicolumn{4}{c}{(\masyr)}
 & \multicolumn{2}{c}{(km s$^{-1}$)} \\   
(\iras1913:                 
 & \multicolumn{2}{c}{\ \hrulefill \ } 
 & \multicolumn{4}{c}{\ \hrulefill \ } 
 & \multicolumn{4}{c}{\ \hrulefill \ } 
 & \multicolumn{2}{c}{\ \hrulefill \ }  \\                                       
I2007 & R.A. & decl. 
 & $\mu^{\mbox{\tiny rel}}_{X}$ & $\sigma \mu^{\mbox{\tiny rel}}_{X}$ 
 & $\mu^{\mbox{\tiny rel}}_{Y}$ & $\sigma \mu^{\mbox{\tiny rel}}_{Y}$
 & $\mu^{\mbox{\tiny abs}}_{X}$ & $\sigma \mu^{\mbox{\tiny abs}}_{X}$ 
 & $\mu^{\mbox{\tiny abs}}_{Y}$ & $\sigma \mu^{\mbox{\tiny abs}}_{Y}$
 
 & V$_{Z}$ & $\Delta$V$_{Z}$ \\ \hline          
  1   \ \dotfill \  &$  -128.19$&$    14.29$&$  -3.44$&   0.02 &$   0.21$&   0.04
&$  -4.10$&   0.02 &$  -3.89$&   0.04
 &$ -53.45$&   1.21 \\
  2   \ \dotfill \  &$  -127.82$&$    13.93$&$  -3.52$&   0.01 &$   0.15$&   0.07
&$  -4.25$&   0.01 &$  -3.93$&   0.07
 &$ -51.10$&   1.15 \\
  3   \ \dotfill \  &$     8.62$&$     9.12$&$  -1.09$&   0.70 &$   0.30$&   0.89
&$  -0.72$&   0.70 &$  -4.18$&   0.89
 &$  44.96$&   1.68 \\
  4   \ \dotfill \  &$    -8.98$&$    -1.62$&$   0.69$&   0.37 &$  -0.08$&   0.41
&$   1.06$&   0.37 &$  -4.56$&   0.41
 &$  47.82$&   0.84 \\
  5   \ \dotfill \  &$   -10.12$&$    -1.66$&$  -0.72$&   0.09 &$  -0.48$&   0.15
&$  -1.70$&   0.09 &$  -4.59$&   0.15
 &$  49.12$&   0.94 \\
  6   \ \dotfill \  &$   -10.68$&$    -1.91$&$   0.34$&   0.18 &$   0.34$&   0.09
&$   0.04$&   0.18 &$  -3.43$&   0.09
 &$  50.85$&   1.90 \\
  7   \ \dotfill \  &$   -10.16$&$    -2.07$&$  -0.01$&   0.33 &$   2.25$&   0.42
&$   0.11$&   0.33 &$  -3.04$&   0.42
 &$  51.20$&   1.47 \\
  8   \ \dotfill \  &$   -11.57$&$    -1.98$&$   0.43$&   0.47 &$  -2.27$&   0.41
&$   0.80$&   0.47 &$  -6.75$&   0.41
 &$  51.35$&   0.84 \\
  9   \ \dotfill \  &$    -9.40$&$    -1.55$&$  -0.73$&   0.05 &$   0.04$&   0.05
&$  -1.50$&   0.05 &$  -4.00$&   0.05
 &$  51.67$&   1.43 \\
 10   \ \dotfill \  &$    -9.67$&$    -1.58$&$  -0.06$&   0.19 &$  -0.10$&   0.37
&$  -0.84$&   0.19 &$  -3.78$&   0.37
 &$  52.04$&   1.47 \\
 11   \ \dotfill \  &$    15.72$&$     5.71$&$  -0.18$&   0.04 &$  -0.28$&   0.05
&$  -1.02$&   0.04 &$  -4.49$&   0.05
 &$  52.17$&   0.70 \\
 12   \ \dotfill \  &$    -8.79$&$    -1.46$&$  -0.14$&   0.07 &$   0.12$&   0.20
&$   0.23$&   0.07 &$  -4.37$&   0.20
 &$  53.03$&   1.38 \\
 13   \ \dotfill \  &$    -8.25$&$    -1.28$&$  -0.88$&   0.01 &$  -0.15$&   0.03
&$  -1.55$&   0.01 &$  -4.22$&   0.03
 &$  53.10$&   0.98 \\
 14   \ \dotfill \  &$    -8.67$&$    -1.67$&$  -0.42$&   0.16 &$   0.06$&   0.19
&$  -0.82$&   0.16 &$  -4.05$&   0.19
 &$  54.16$&   0.84 \\
 15   \ \dotfill \  &$    11.94$&$     7.70$&$  -0.08$&   0.07 &$  -0.34$&   0.10
&$  -0.54$&   0.07 &$  -4.32$&   0.10
 &$  54.67$&   0.47 \\
 16   \ \dotfill \  &$     0.00$&$     0.00$&$   0.00$&   0.01 &$   0.00$&   0.02
&$  -0.68$&   0.01 &$  -4.10$&   0.02
 &$  54.71$&   0.94 \\
 17   \ \dotfill \  &$    -3.96$&$    -0.97$&$   0.25$&   0.27 &$   2.18$&   0.37
&$   0.38$&   0.27 &$  -3.11$&   0.37
 &$  55.75$&   0.47 \\
 18   \ \dotfill \  &$    -4.34$&$    -0.88$&$   0.14$&   0.04 &$  -0.11$&   0.08
&$  -0.36$&   0.04 &$  -4.12$&   0.08
 &$  55.76$&   1.05 \\
 19   \ \dotfill \  &$    -2.15$&$    -0.24$&$  -0.68$&   0.02 &$  -0.24$&   0.02
&$  -1.36$&   0.02 &$  -4.42$&   0.02
 &$  55.89$&   1.58 \\
 20   \ \dotfill \  &$    -4.63$&$    -0.98$&$  -0.20$&   0.18 &$   0.01$&   0.09
&$  -0.75$&   0.18 &$  -3.94$&   0.09
 &$  56.44$&   0.91 \\     
 \hline
 \end{tabular}

\tablenotetext{1}{Maser features detected in \iras1913. 
The features are designated as \iras1913:I2007 {\it N}, where {\it N} is the ordinal 
source number given in this column (I2007 stands for sources found by 
Imai et al. and listed in 2007).}
\tablenotetext{2}{Relative value with respect to the location of 
the position-reference maser feature: \iras1913:I2007 {\it 16}.}
\tablenotetext{3}{Relative value with respect to the motion of the position-reference 
maser feature: \iras1913:I2007 {\it 16}.}
\tablenotetext{4}{Relative value with respect to the adopted systemic velocity 
$V_{\mbox{\tiny LSR}}=-67$\kms.}

\end{table*}

\section{Results}

\subsection{The morphology of optical lobes and the maser spatio-kinematics in \iras1913}
\label{sec:kinematics}

\begin{figure*}
\epsscale{1.5}
\plotone{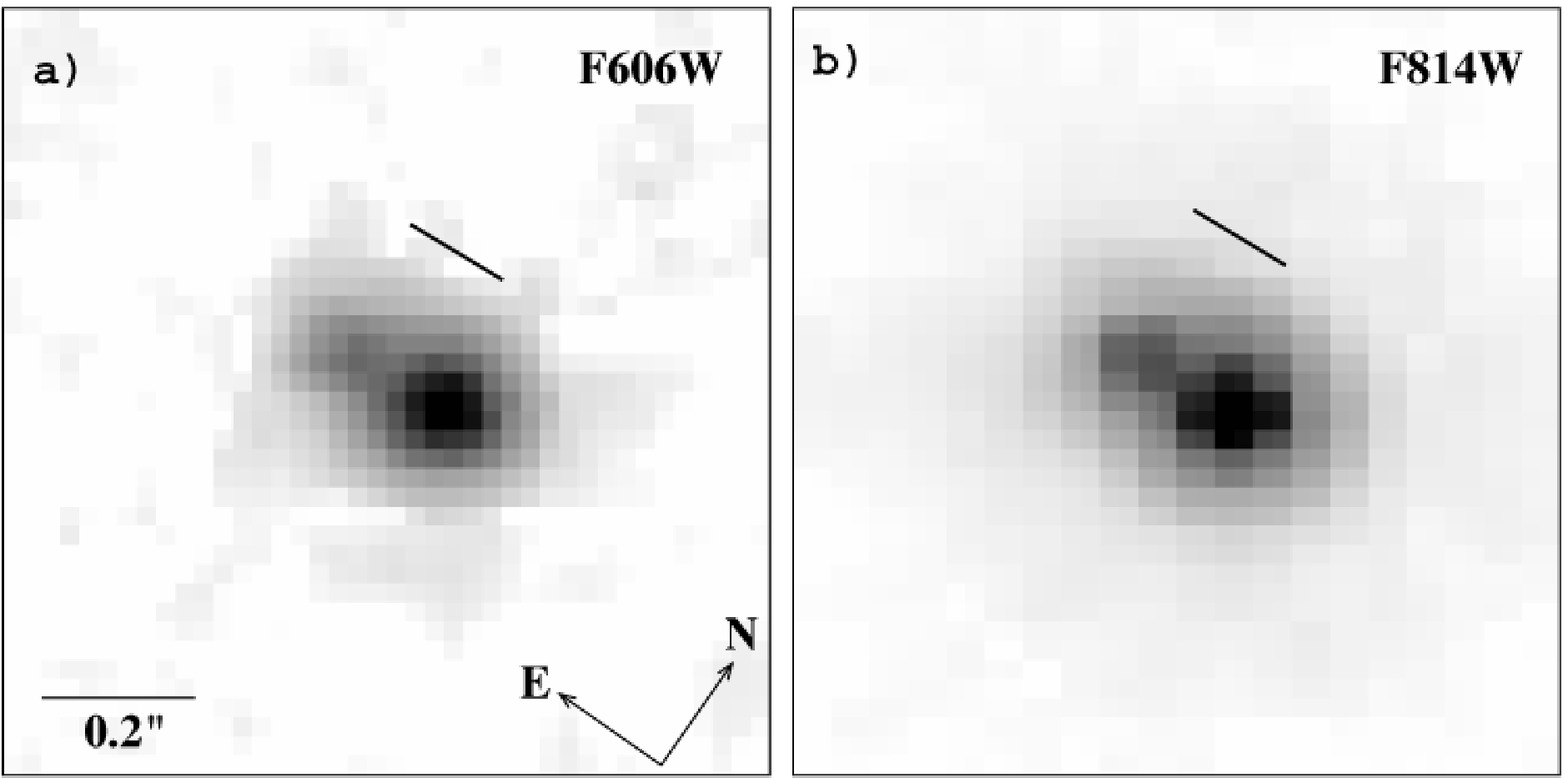}
\caption{Image of the pre-planetary nebula \iras1913, made using exposures taken with ACS/HST through the (a) F606W filter, and (b) F814W filter. The angular scale and orientation are identical in both panels. Images show the square-root of the intensity (on a linear stretch). The maximum ({\it black}) and minimum ({\it white}) intensities of the grey scale are $7.0\times 10^{-17}$ and  
$2.3\times 10^{-18}$~erg~s$^{-1}$~cm$^{-2}$ \AA$^{-1}$~arcsec$^{-2}$, 
respectively, in panel (a), and $4.5\times 10^{-16}$ and $1.1\times 
10^{-17}$~erg~s$^{-1}$~cm$^{-2}$ \AA$^{-1}$~arcsec$^{-2}$, respectively, in panel 
(b). The solid line above the bipolar nebulosity shows the position angle 
($PA=$94\arcdeg) of the projected flow expansion vector derived from the \h2o 
masers; its size (139~mas) is equal to the separation of the red and blue-shifted maser features.
\label{fig:HST}}

\ \\
\epsscale{1.5}
\plotone{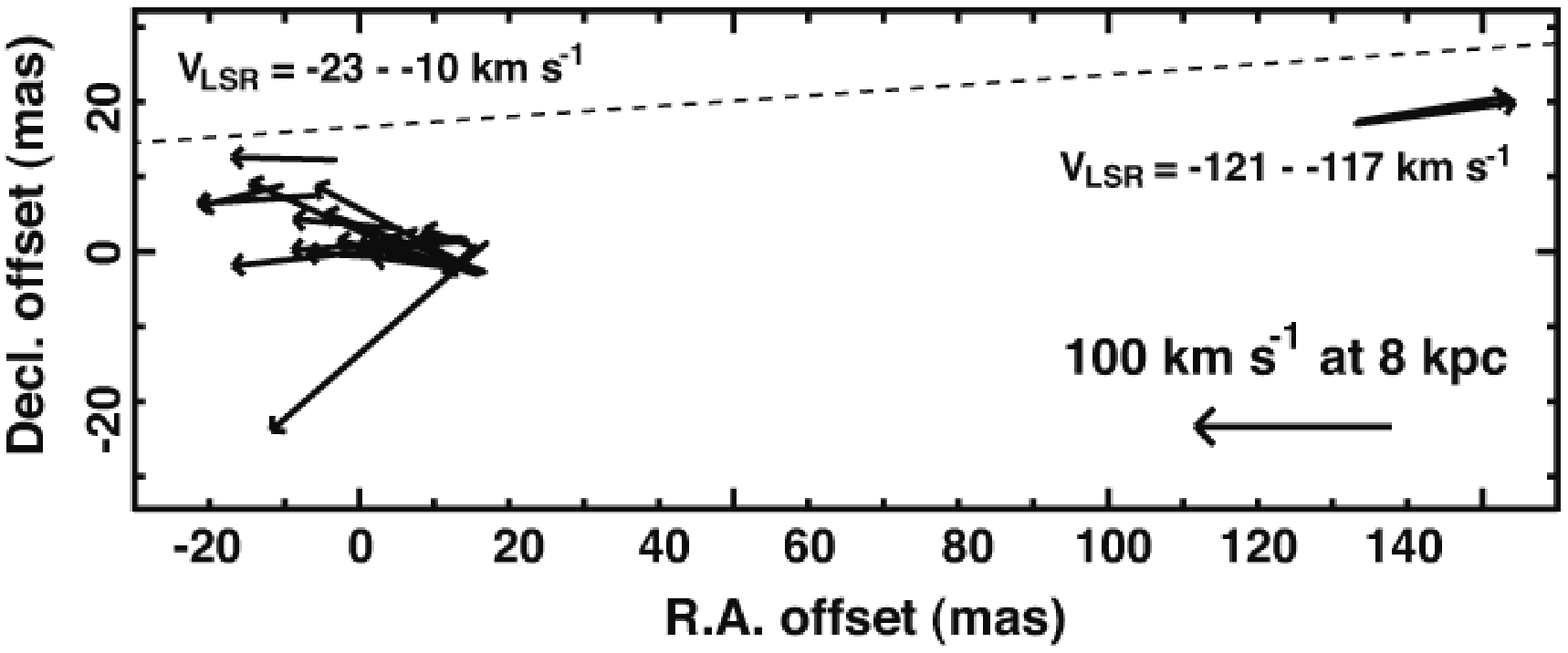}
\caption{Relative proper motions of \h2o\ masers in \iras1913, with the systemic motion subtracted. Position offsets with respect to the location of the position-reference maser feature: \iras1913:I2007 {\it 16} are shown here. A thin dashed line indicates the direction of the bipolar flow estimated from the proper motion vectors.
\label{fig:RPM}}
\end{figure*}

Figure \ref{fig:HST} shows HST optical images of I19134. There exist two optical lobes separated by $ \sim$150~mas, whose coordinates are estimated using seven stars that are catalogued in the USNO B.1 catalogue and located within 30\arcsec\ from I19134.  The intensity peak of the western lobe is located at 
$\alpha_{J2000}=$19$^{h}$15$^{m}$35$^{s}$\hspace{-2pt}.20,
$\delta_{J2000}=$~$+$21$^{\circ}$36$^{\prime}$34\farcs 5. The 1-$\sigma$ uncertainty in the coordinates is 0\arcsec.33 in both R.A.\ and decl. The centroid of the eastern lobe is located about 0\arcsec.13 east of this location in the F606W image.  There is no ambiguity about which star to associate with the central object of the \h2o masers because this is the only bright red object visible in the 2MASS near-infrared bands within 2\arcsec\ of the masers. The difference in the locations of the centroids of the maser feature groups and the optical lobes is only about 0\farcs 15 in R.A., but about 0\farcs 6 in declination, which is consistent with being coincident, within 2-sigma. However, the 2MASS coordinates are coincident with the centroid of the maser coordinates within 0\farcs 1. 

Figure \ref{fig:RPM} shows the distribution of \h2o maser features. The maser emission consists of two clusters of maser features; the blue-shifted (\vlsr$=-$121 to $-$117\kms) and red-shifted (\vlsr$=-$23 to $-$10\kms) maser components are clearly distinguished and are separated by $ \sim$140~mas. As shown in figure \ref{fig:HST}, the optical emission is elongated in the same direction as the major axis of the \h2o maser distribution. The western optical component is brighter and may be closer to us than the eastern fainter component, consistent with the flow inclination found in the maser emission with respect to the sky plane. The large brightness ratio of the two optical lobes suggests that the central evolved star is located at the middle of the lobes and still has a thick CSE that significantly obscures optical emission from the eastern lobe. 

We measured {\it relative} proper motions of the \h2o\ maser features with respect to the position reference feature \iras1913:I2007 {\it 16}. Columns 4--7 in table \ref{tab:pmotions} list, for 20 maser features that were detected twice or more, their measured proper motions. The mean motions of the red-shifted and blue-shifted feature clusters were measured, then a mean systemic motion of the I19134 \h2o\ maser features was derived from the mean of the means of the two cluster motions. Thus we obtain a {\it relative} mean systemic motion of 
$(\Delta \mu_{X}, \Delta \mu_{Y})=
(-1.84\pm0.81, 0.13\pm0.76)$ [\masyr] 
and a one-way projected flow expansion vector of $1.85\pm0.63$\masyr\ 
with P.A.$=$94\arcdeg. Here $X$ and $Y$ denote the east--west and north--south directions, respectively. Figure \ref{fig:RPM} shows the proper motion vectors of maser features with the systemic motion subtracted. The systemic radial velocity of $V_{\mbox{\tiny LSR}}=-67\pm2$\kms\ is also obtained by applying the same procedure to the mean radial velocity of each of the two clusters. The resulting three-dimensional expansion velocity for each cluster, with respect to the mean systemic motion, is 89$\pm$25\kms\, at an inclination of $35\arcdeg^{+13\arcdeg}_{-7\arcdeg}$ with respect to the sky plane. Here we adopt a distance to I19134 to be 8~kpc (see the next section). 

\begin{table}[ht]
\tablenum{4}
\caption{Diagonalization analysis of the variance-covariance matrix of 
the velocity and position vectors of the \iras1913 \h2o\  masers.}
\label{tab:vvcm}
\begin{tabular}{ccc} \hline \hline
& \multicolumn{2}{c}{Eigenvector} \\
Eigenvalue & Inclination & Position angle \\ \hline
\multicolumn{3}{c}{Velocity variance-covariance matrix} \\
$\left[(\mbox{km~s}^{-1})^{2}\right]$ & & \\ \hline
$2754\pm88$ & 36\arcdeg.0$\pm$3\arcdeg.3 & 92\arcdeg.0$\pm$8\arcdeg.4 \\
$1253\pm200$ & 1\arcdeg.0$\pm$4\arcdeg.4 & 1\arcdeg.8$\pm$3\arcdeg.0 \\ 
$129\pm34$ & 53\arcdeg.5$\pm$1\arcdeg.1 & $-$89\arcdeg.5$\pm$5\arcdeg.1 \\
\hline
\multicolumn{3}{c}{Spatial variance-covariance matrix} \\ 
(mas$^{2}$) & & \\ \hline
1526 & ... & 95\arcdeg.4 \\ 
16 & ... & 5\arcdeg.4 \\ \hline
\end{tabular}

\end{table}

In order to find the axis of the flow more objectively and precisely, we performed diagonalization for the velocity variance--covariance matrix obtained from velocity vectors of maser features (c.f., \citealt{blo00}). The eigenvector corresponding to the largest eigenvalue (velocity dispersion) indicates the major axis of the flow. A ratio of square roots of the largest and second largest eigenvalues gives a collimation factor of the flow.  The uncertainties in the obtained eigenvectors and eigenvalues were derived from standard deviations of these parameters, which were calculated with a Monte--Carlo simulation for the matrix diagonalization using velocity vectors varying around the observed values within their estimated errors. Table \ref{tab:vvcm} gives the eigenvalues and their corresponding eigenvectors obtained after the matrix diagonalization. The spatial eigenvector is also parallel to the flow axis within the uncertainty of their directions ($ \sim$8\arcdeg). The collimation factors of the maser velocity vectors and the positional maser feature distributions are 4.8 and 9.7, respectively. These factors are smaller than those of the water fountain in W43A \citep{ima02}, but still indicate kinematically and spatially high collimation of the flow, similar to W43A. 

We note that there is an aligned distribution of maser spots in the red-shifted maser cluster having an angle offset of $\sim$10\arcdeg\ from the flow axis (figure \ref{fig:RPM} and \ref{fig:PM}). This slant of the aligned maser distribution may indicate the presence of a precessional motion in the flow. Alternatively, the masers might be distributed only on one side of a bow shock for some reason, so that the tilted distribution of spots just traces half of the bow shock. If the maser proper motion vectors are parallel to the flow axis rather than to the direction of elongation of the maser distribution, the former is possible. However, the observed proper motion vectors show a significant scatter ($\sigma=$22\arcdeg) around the mean value of P.A.$=$94\arcdeg. Investigation of proper motion alignment for a much longer time span such as that carried out for W43A (INDMD05) is necessary to unambiguously obtain a reliable spatio-kinematical model for the \h2o\ masers and to assess the precessing jet hypothesis. Such long-term observations are also necessary to estimate a terminal jet velocity. \citet{bob05} found an acceleration motion in the \h2o\ maser spectra of OH~12.8$-$0.9 at a rate, 0.68$\pm$0.06\kms~yr$^{-1}$, during a period of 20~yr. We do not find systematic acceleration/deceleration motions of the individual maser features, with an upper limit of 1.5\kms~yr$^{-1}$. Because the maser features most likely represent jet-compressed regions of the AGB envelope rather than the jet material itself (e.g., \citealt{eli89}), it is difficult to find any change in the intrinsic jet speed during the lifetime of a single maser feature. 

The dynamical age of the flow is estimated to be $38\pm11$~yr, which is roughly equal to that of W43A ($\sim$50~yr, INDMD05). Both were estimated on the basis of maser proper motions. Note that a proper motion speed is roughly equal to the linear growth rate of the jet and that such an estimation method is expected to provide the true time scale of the jet evolution since the jet's initiation (INDMD05). For other water fountains, similar dynamical ages ($\lesssim$100~yr) have been found  \citep{mor03,bob05}. Thus we have found a lifetime that is rather typical of the water fountains, although it should be more carefully determined with a larger sample of objects in this class. 

\subsection{Astrometry}
\label{sec:astrometry}

\begin{figure*}
\epsscale{2.0}
\plotone{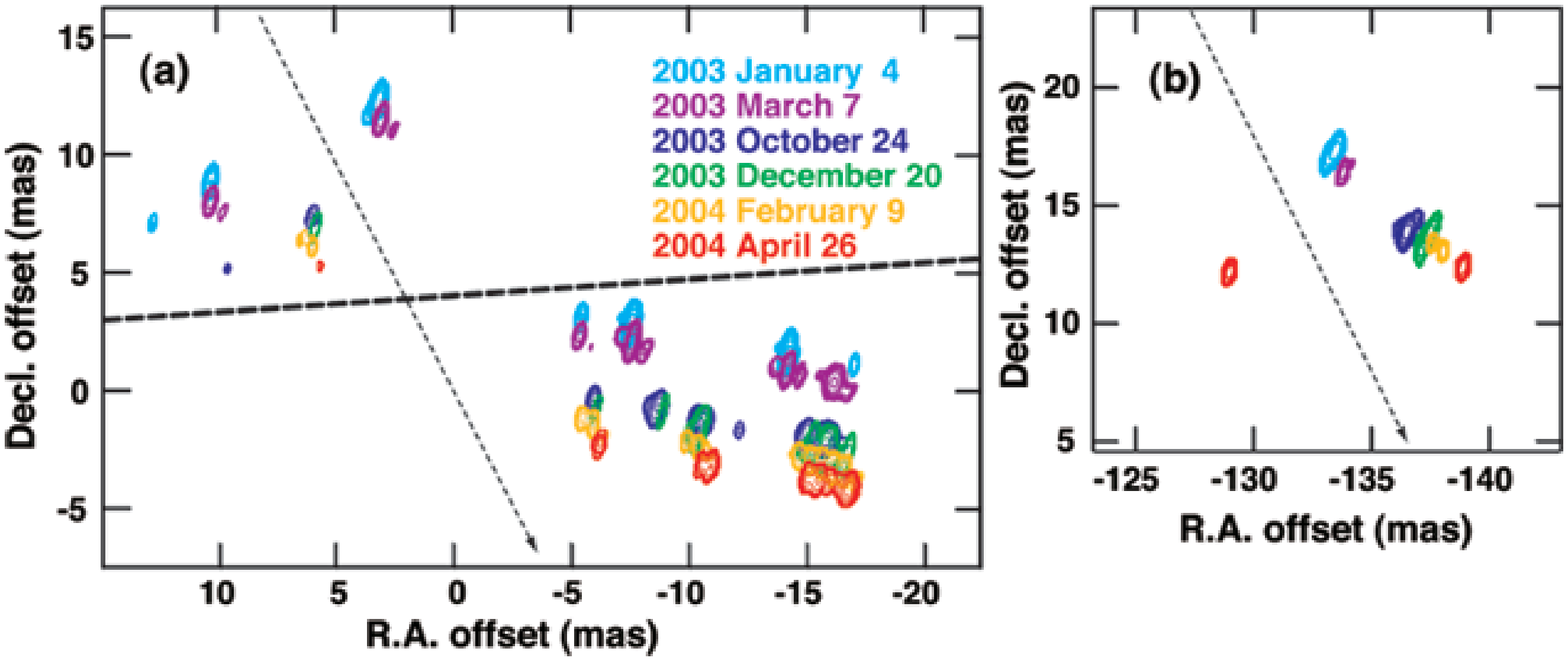}
\caption{Temporal variation of the spatial distribution of \h2o\ maser emission 
in \iras1913. Position offsets with respect to the delay-tracking center at data correlation (see the main text) are shown here. Contour levels are logarithmically set with respect to the largest intensity in each map and at each epoch. The thick dashed line indicates the direction of the flow axis. The thin dashed arrow indicates the direction of Galactic rotation. The red-shifted ({\it a}) and blue-shifted ({\it b}) clusters of maser features are shown on a finer scale than in Fig.\ 2.
\label{fig:PM}}

\ \\
\epsscale{2.0}
\plotone{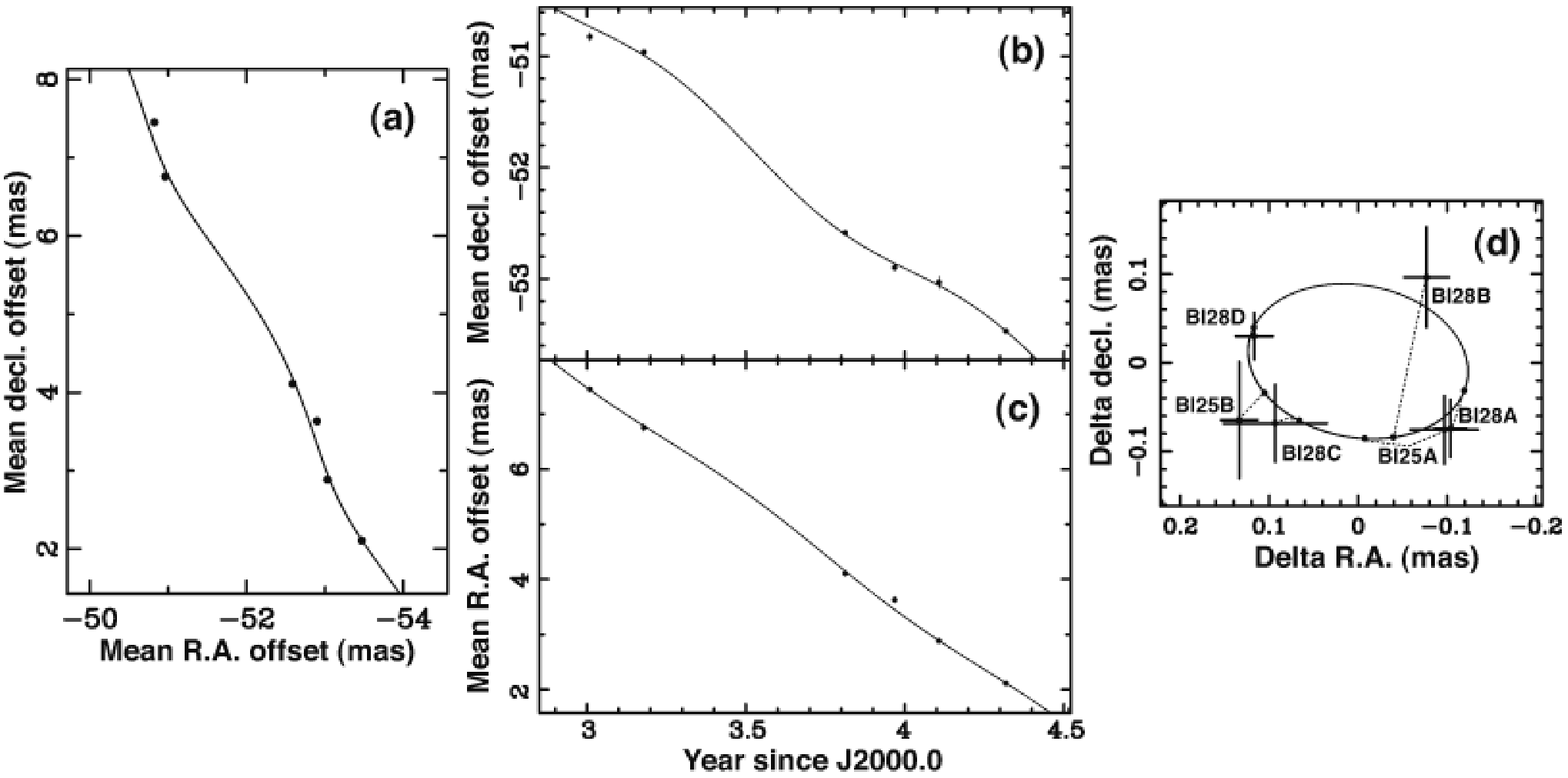}
\caption{Mean position of the three maser features \iras1913:I2007 {\it 1}, {\it 13}, and {\it 16} at all of the observation epochs. ({\it a}) Mean R.A.\ and decl.\ offsets on the sky with respect to the phase-tracking center. ({\it b}) and ({\it c}) Mean R.A.\ and decl.\ offsets against time. ({\it d}) Relative mean offsets with a position offset on the date J2000.0 and a mean proper motion subtracted. The observation code is shown alongside each measured position. An ellipse indicates the modeled annual-parallax motion. The mean maser position goes clockwise around the ellipse. 
The observed point is connected to the relevant point on the ellipse, corresponding to the observation time, with a dotted line. 
\label{fig:parallax}}
\end{figure*}

In \S \ref{sec:kinematics}, {\it relative} maser proper motions are discussed, in which they are measured with respect to one of the maser features. In this section, {\it absolute} maser proper motions, or changes of maser feature positions measured with respect to the reference, J1925, are presented. Figure \ref{fig:PM} shows the temporal variation of the velocity-integrated maser intensity distribution in I19134, in the coordinate system fixed with respect to the position reference, J1925. The blue-shifted and red-shifted clusters of maser features are moving in the SSW and SW directions, respectively; the difference corresponds to the separating motion described in \S \ref{sec:kinematics}. In addition, a systemic secular motion toward the Galactic center (in the SSW direction) is clearly seen. Columns 8--11 in table \ref{tab:pmotions} show the {\it absolute} maser proper motions. From these values, and weighting them by the inverses of the proper motion errors, we obtain the systemic secular motion of I19134. Here we assume that the red-shifted and blue-shifted  
clusters are moving away from the central source at the same velocity but in  opposite directions.  This is likely to be a good assumption, given the velocity symmetry shown by this class of sources \citep{lik92}. Thus we obtain the systemic secular motion in the east--west ($X$) and north--south ($Y$) directions: 

\begin{equation}
(\mu_{X}, \mu_{Y})=(-2.35\pm0.81, -4.05\pm0.76)[\mbox{mas~yr$^{-1}$}].  
\end{equation}

\noindent
In Galactic coordinates, this becomes:
\begin{equation}
(\mu_{l}, \mu_{b})=(-4.7\pm1.0, 0.3\pm1.1)[\mbox{mas~yr$^{-1}$}].  
\end{equation}

\noindent
Furthermore, we find three maser features which were detected at all six epochs and are confirmed to exhibit a common annual parallax -- \iras1913:I2007 {\it 1}, {\it 13}, and {\it 16}.  We therefore fit the temporal variation of the mean position of the three features to a kinematical model consisting of an annual parallax, $\pi$, and a constant velocity mean secular motion, $(\mu_{X}, \mu_{Y})$:

\begin{eqnarray}
\pi & = & 0.125\pm0.012 \mbox{ mas} \\
\mu_{X} & = & -2.186\pm0.022 \mbox{\masyr} \\
\mu_{Y} & = & -4.165\pm0.033 \mbox{\masyr} \\
\Delta X_{2003.01} & = & 6.677\pm0.084 \mbox{ mas} \\ 
\Delta Y_{2003.01} & = & 12.612\pm0.130 \mbox{ mas},
\end{eqnarray}

\noindent
where $\Delta X_{2003.01}$ and $\Delta Y_{2003.01}$ are the position offsets with respect to the delay-tracking center on the date 2003.01. Figure \ref{fig:parallax} shows the fitting result. The mean motion of the three features are {\it roughly} consistent with the kinematical model including the annual parallax. The mean position should go clockwise around the annual-parallax ellipse. However, the positions at BI25A and BI28B are 150~$\mu$as west and 180~$\mu$as north, respectively, from those expected from the modeled parallax (see figure \ref{fig:parallax}{\it d}). Even with such large position offsets, the fit to the annual parallax turns out to be quite reasonable. The best-fit annual parallax corresponds to a distance to I19134, $D=8.0^{+0.9}_{-0.7}$~kpc. 

\section{Discussion}
\subsection{The water fountain in \iras1913}
\label{sec:fountain}

The implications of the water fountains for the formation of bipolar PNe were discussed in Paper {\rm I}, but the new observations presented here should provide new insights into the general properties of the water fountains and their associated central stars.

The spatio-kinematics of \h2o masers in all water fountain sources reveal that the duration of the water-fountain stage is less than 100~yr. In this time interval, a stellar jet traveling with a typical velocity of 100\kms\ reaches a distance of $ \sim$2000~AU from the central star. This corresponds to the typical size of an OH maser shell in a CSE around an OH/IR star, at the final stage of copious mass loss. The \h2o maser emission is excited near the tip of the stellar jet where high compression takes place in the shock at the interface between the jet and the ambient circumstellar gas. But because the ambient gas density declines with radius, the maser emission may be quenched at the distance where the post-shock gas density falls below a critical value, $n_{\mbox{\tiny  H$_{2}$}}\sim$10$^6$ cm$^{-3}$ \citep{eli89}. 
Note that while optical lobes are observed in some water fountains (I19134 and IRAS~16342$-$3814, hereafter abbreviated as I16342), the optical emission has not yet been detected in others (W43A and OH~12.8$-$0.9).  This implies that the presence of water fountains may help in the identification of new PPNe, which may be optically invisible due to high circumstellar or interstellar extinction. On the other hand, the number of water fountains has grown by almost a factor of four since this class was first recognized \citep{lik88,lik92}, implying that the lifetime of each water fountain may be much longer than that previously estimated (e.g., \citealt{ima02}). 

Alternatively, the water fountain phenomenon may occur repeatedly in the same source. However, it is premature to give a conclusion about this issue. The total number of stellar \h2o masers associated with AGB and post-AGB stars is still increasing and is too small ($<$1,000, e.g., \citealt{val01,tak01}) to precisely estimate the duration of the water fountain phase whose number of occurrences is only about 10. If the water fountain lifetime estimated here is presently known, the total number of stellar \h2o masers in the whole AGB phase gives the duration of \h2o maser excitation, which is inversely estimated from the lifetime of the water fountains ($ \sim$100~yr). The duration of \h2o maser excitation is thereby estimated to be only $<$10$^{4}$~yr. This is much shorter than $\sim 10^{5}$~yr, which is the expected time period for excitation of \h2o masers during AGB/post-AGB phase. On this time scale, an AGB star loses a total mass of $ \sim$1~$M_{\sun}$ at the maximum stellar mass loss rate (typically 10$^{-5}M_{\sun}$~yr$^{-1}$), which is sufficiently high for excitation of \h2o masers (e.g., \citealt{bow94}). Note that not all stellar \h2o maser sources necessarily pass through the water fountain stage; bipolar nebulae that have jets may be a restricted class of stellar \h2o masers. 

As shown in \S \ref{sec:kinematics}, the optical lobes in I19134 coexist with the \h2o maser emission. This is a case similar to that found in I16342, where the masers lie just beyond the tips of the optical lobes \citep{sah99,mor03}. We adopt the paradigm that the lobes have recently been formed along with the collimated fast jet. Then the CSE around the evolved star may have been sculpted and evacuated by the jet, allowing starlight to escape and be scattered by dust in, or at the edge of, the resulting lobes. I19134 and I16342 are presumably at a similar transitional stage between the AGB and post-AGB phases. Also the inclination of the I16342 jet is estimated to be $ \sim$40\arcdeg\ with respect to the sky plane \citep{sah99}, similar to that of I19134 ($\sim$35\arcdeg). The ratios of peak surface brightness between the two lobes are 3.4 and 6.2 for I19134 and I16342, respectively, but the low spatial resolution in I19134 might reduce the ratio slightly from the intrinsic contrast. I16342 has larger separations of the optical lobes and the maser components (1\farcs 8 and 3\arcsec\ or $\sim$3600~AU and 6000~AU at 2~kpc, respectively) than those in I19134 ($\sim$1200~AU). However, the difference in the dynamical ages of the jets in these two systems is still comparable to their implied dynamical ages ($\leq$100~yr). 

Does the size of a CSE determine the lifetime of a water fountain, or a collimated stellar jet traced by molecular emission?  We already know that \h2o masers can survive until photoionization of the CSE starts to form a PN, such as in K3$-$35 and IRAS~17347$-$3139  \citep{mir01, gre04}, but they do not exhibit water fountains. In K3$-$35, \h2o maser emission is seen at the tips of the radio lobes, but it is not morphologically collimated so much and the radial velocities of the masers are very slow (a radial velocity range of only 15\kms). This implies that any initially collimated jet that might have been present and traceable by \h2o maser emission has disappeared. In K3$-$35, \h2o maser emission is also seen in the inner and denser part of the CSE where UV radiation from the star has not yet penetrated. This raises the possibility that \h2o maser emission can be excited while the spatio-kinematical  collimation of a water fountain is being destroyed. Such low velocity maser components have been found in other water fountains alongside the high velocity ones \citep{ima07}. The dynamical age of the radio lobes of K3$-$35 is estimated to be 800~yr \citep{mir01}. This may correspond to an upper limit to the lifetime of a water fountain source but is an order of magnitude longer than the dynamical age estimated using VLBI observations of the maser spatio-kinematical structures. In I19134, the morphological similarity of the two broad-band images shown in figure \ref{fig:HST}  suggests that the optical lobes are due to scattered starlight, rather than emission lines as in a PN. Therefore we speculate that the collimated jet may drill through the CSE formed during the AGB phase and the tip of the jet may soon reach the outer boundary of the envelope. There the density declines precipitously and the maser levels are no longer sufficiently excited to produce observable emission.  The masers lying beyond the optical lobes, such as those in I16342, might thus disperse and disappear within a few centuries owing to the decline of the gas density in the pre-shock gas as the jet propagates outward. Does the CSE size play a significant role in determining the lifetime of the water fountain? The question can be answered with a statistical comparison of the dynamical time scales for the water fountains to the time scale on which photoionized radio lobes emerge and expand along the axis of the water fountains. 

\subsection{The location and kinematics of \iras1913 in the Galaxy}
\label{sec:distance}

\begin{figure}
\epsscale{1.0}
\plotone{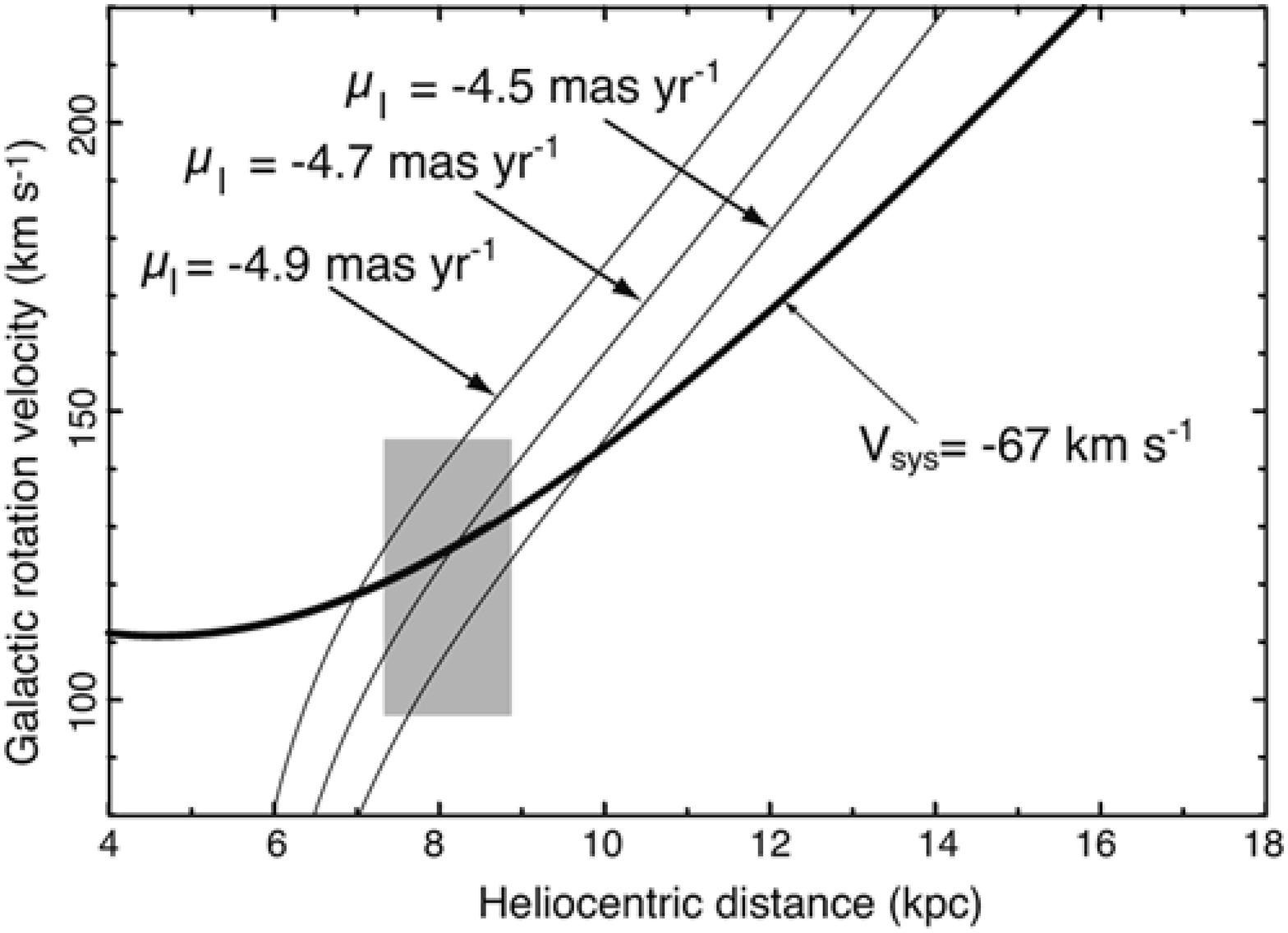}
\caption{Possible heliocentric distance, $r$, and Galactic azimuthal velocity, 
$V_{\theta}$, of I19134 located at $(l, b)=$(54\arcdeg.867, 4\arcdeg.642). 
Three curves of the function corresponding to the proper motions, $\mu_{l}\:=$~$-$4.5, $-$4.7, $-$4.9\masyr, respectively, are plotted together with another bold curve corresponding to the source radial velocity \vlsr$=$~$-$67\kms. The latter curve assumes that the source velocity vector is perpendicular to the radius to the Galactic center. The grey box indicates the possible range of the heliocentric (parallactic) distance and the Galactic azimuthal velocity of I19134 obtained from the three-dimensional velocity vector.  
\label{fig:kinematic-distance}}
\end{figure}

In the present analysis, we adopt a galactocentric distance and a Galactic rotation velocity of the Sun to be 8.0~kpc and 217\kms, respectively \citep{rei93, deh98}. Then, on the basis of our inferred heliocentric distance and the systemic secular motion of I19134, we infer the location of I19134 [$(l,b)=$(54\arcdeg.867, 4\arcdeg.642)] in the Galaxy to be, 
\begin{equation}
(R, \theta, z)=(7.4^{+0.4}_{-0.3}\mbox{ kpc}, 
62\arcdeg\pm 5\arcdeg, 0.65^{+0.07}_{-0.06}\mbox{ kpc}),  
\end{equation}
\noindent
and the 3-D velocity vector in Galactic cylindrical coordinates to be, 
\begin{equation}
(V_{R}, V_{\theta}, V_{z})=
(3^{+53}_{-46}, 125^{+20}_{-28}, 8^{+48}_{-39})\mbox{ [km~s$^{-1}$]}.
\end{equation} 

\noindent 
The bias due to the Solar motion with respect to the local-standard of rest has been corrected, but is relatively small ($\sim$3\kms). The derived heliocentric and galactocentric distances of 8.0~kpc and 7.4~kpc, respectively, are much smaller than those ($\sim$16~kpc and $\sim$13~kpc, respectively) estimated by the kinematical distance method using only the systemic radial velocity (\vlsr$=-$67\kms). As described in Paper {\rm I} and shown in figure \ref{fig:kinematic-distance}, the kinematical distance obtained using both the radial velocity and the secular proper motion, with the assumption of circular Galactic rotation, is estimated to be about 8.0~kpc. This is consistent with the annual-parallax distance. The estimated location is within 1~kpc of the Perseus spiral arm (c.f., \citealt{nak06}). 

Note, however, that the estimated Galactic azimuthal velocity of I19134 ($\sim$125\kms) 
is much slower than that implied by the Galactic rotation curve ($\sim$220\kms, e.g., \citealt{cle85, deh98}). It implies that I19134's orbit must have large excursions in radius although it must be orbiting near its maximum in galactocentric distance at present. I19134 is apparently orbiting as a member of the Galactic ``thick disk`` component rather than the ``thin disk`` component. Thus the high Galactic latitude of I19134 favors a low mass of I19134, allowing for it to orbit as a member of the Galactic thick disk component. Adopting the source velocity perpendicular to the Galactic plane, $V_{z}=$8\kms, we find that the present Galactic altitude gives a lower limit to the travel time of I19134 from the Galactic plane to the present position to be $\sim 2.4\times$10$^{7}$~yr. This suggests that the central star of I19134 has a lifetime of at least this travel time and a stellar mass smaller than  $\sim$ 8.6~$M_{\sun}$ (here we adopt a stellar lifetime to be $t_{\ast}\propto M_{\ast}/L_{\ast}\propto M^{-2.8}_{\ast}$) if the central star of I19134 was born in the Galactic plane as a member of the Galactic ``thin disk`` component. The timescale for a star that initially has thin disk kinematics to acquire thick disk kinematical characteristics is very likely to be far larger than 2.4$ \times$10$^{7}$~yr, so I19134 is therefore likely to have a much larger age than this timescale. The statistical study of Galactic PNe by \citet{man04} shows that the Galactic scale height of bipolar PNe, such as I19134 may become, is $<z>=$100~pc, suggesting that their progenitors should be high mass stars. Although I19134's progenitor is expected to be such a high mass star, the present Galactic altitude of I19134 and its anomalously slow Galactic rotation velocity are quite odd. These may imply an abnormal environment during the birth of I19134's central star. Thus, the mass of I19134's progenitor is an interesting puzzle.  

\section{Conclusions}

Together with the spatio-kinematical structures of other water fountain sources, 
the spatio-kinematical structures of \h2o masers in I19134 have shown that the dynamical ages of the water fountains are shorter than 100~yr. More precise ages and more detailed physical conditions of such stellar jets should be elucidated by future observations of thermal molecular emission in infrared and radio wavelengths with high angular resolution ($<<$0\farcs 1) to trace the whole morphology of the jets, including the regions without maser emission. 

Optically visible lobes resulting from scattered starlight are found in I16342 and I19134 but have not yet been found in W43A and other water fountain sources. This implies that a water fountain is produced during the transition from the AGB phase, when mass loss is greatest and the CSE is the thickest, to the post-AGB phase. The jet carves bipolar cavities inside the CSE so that it is observed as a bipolar PPN. Subsequently, as the central star evolves to a temperature of about 30,000~K, photoionization sets in and transforms it into a PN.

High precision astrometry for \h2o masers at the 10-$\mu$as level has enabled us to directly measure the distances to a Galactic source located up to 8~kpc from the Sun and to determine the location and kinematics of such distant sources throughout the Galaxy (e.g., \citealt{hon07}).  These parameters of the water fountain sources may provide additional probes to estimate ages and masses of their central stars, helping us to understand which progenitor stellar mass range produces the water fountain sources. For I19134, we obtained an upper limit to the progenitor mass, $M_{\ast}\lesssim 8.6$~$M_{\sun}$. The accuracy of the proper motion of and parallactic distance to I19134, and thus its full three-dimensional velocity found after the intrinsic maser motions are subtracted, may be improved with sensitive VLBA observations over a longer time baseline.  To find a better kinematical model for the intrinsic motions, in particular, point symmetry of the maser spatio-kinematics should be examined. 

\acknowledgments
We acknowledge the referee for carefully reading our paper and giving us many critical and fruitful comments. NASA/ESA HST is operated by the Association of Universities for Research in Astronomy.  The VLBA/National Radio Astronomy Observatory is a facility of the National Science Foundation, operated under a cooperative agreement by Associated Universities, Inc. HI has been financially supported by Grant-in-Aid for Young Scientists (B) from the Ministry of Education, Culture, Sports, Science, and Technology (18740109). RS and MM thank NASA for partially funding this work by a NASA LTSA award (no. 399-30-61-00-00); RS has also received funding from HST/GO awards (no. GO-09463.01-A and GO-09801.01-A) from the Space Telescope Science Institute (operated by the Association of Universities for Research in Astronomy, under NASA contract NAS5-26555).


\begin{deluxetable}{@{}r@{ }r@{ $\pm$}r@{  }r@{ $\pm$}rr@{ }r@{ }c}
\rotate
\tablenum{2}
\tablecaption
{Parameters of the detected \h2o\ maser features. \label{tab:features}}
\tabletypesize{\scriptsize}
\tablecolumns{8}
\tablehead{$V_{\mbox{\tiny LSR}}$\tablenotemark{1} 
& \multicolumn{2}{c}{R.A. offset\tablenotemark{2}} & 
\multicolumn{2}{c}{Decl. offset\tablenotemark{2}} & 
$I$\tablenotemark{3}   & \multicolumn{1}{c}{$\Delta V$\tablenotemark{4}} 
& F\#\tablenotemark{5}}
\startdata
\multicolumn{8}{c}{03/01/04} \\ \hline
  $-$10.99 &  $-$7.626 & 0.011 &   3.844 & 0.014 &     0.23 &   1.05 \\
  $-$11.05 &  $-$6.769 & 0.028 &   4.361 & 0.054 &     0.13 &   0.42 \\
  $-$11.11 &  $-$7.495 & 0.005 &   2.875 & 0.009 &     0.89 &   2.11 & 19 \\
  $-$11.12 &  $-$7.269 & 0.011 &   1.864 & 0.038 &     0.27 &   1.05 \\
  $-$11.20 &  $-$6.974 & 0.017 &   3.194 & 0.039 &     0.16 &   0.84 \\
  $-$11.57 &  $-$7.630 & 0.020 &   3.950 & 0.020 &     0.09 &   0.21 \\
  $-$12.29 &  $-$5.344 & 0.014 &   3.115 & 0.024 &     0.15 &   0.84 & 16 \\
  $-$13.97 &    $-$14.135 & 0.001 &   1.657 & 0.015 &     0.61 &   1.48 & 12 \\
  $-$14.02 &    $-$14.287 & 0.030 &   2.637 & 0.016 &     0.13 &   0.84 \\
  $-$14.02 &    $-$13.914 & 0.017 &   0.626 & 0.055 &     0.15 &   0.84 \\
  $-$13.89 &  14.310 & 0.030 &   6.060 & 0.080 &     0.10 &   0.21 \\
  $-$13.89 &    $-$13.900 & 0.011 &   0.440 & 0.030 &     0.13 &   0.21 \\
  $-$13.90 &    $-$13.590 & 0.011 &   1.832 & 0.047 &     0.12 &   0.63 & 13 \\
  $-$14.83 &  10.373 & 0.027 &   8.825 & 0.016 &     0.23 &   0.84 & 11 \\
  $-$15.65 &    $-$16.918 & 0.028 &   1.132 & 0.041 &     0.09 &   0.63 & 8 \\
  $-$16.85 &  12.848 & 0.030 &   7.078 & 0.061 &     0.08 &   0.63 \\
  $-$19.18 &    $-$14.325 & 0.053 &   1.491 & 0.043 &     0.12 &   1.05 & 4 \\
  $-$20.63 &   3.850 & 0.020 &  11.620 & 0.040 &     0.09 &   0.21 \\
  $-$22.04 &   3.279 & 0.103 &  12.235 & 0.118 &     0.23 &   1.90 & 3 \\
 $-$118.10 &   $-$133.160 & 0.010 &  17.040 & 0.084 &     0.13 &   1.68 & 2 \\
 $-$120.45 &   $-$133.534 & 0.063 &  17.408 & 0.040 &     0.20 &   1.90 & 1 \\
 \hline
\multicolumn{8}{c}{03/03/07} \\ \hline
  $-$11.10 &  $-$7.050 & 0.046 &   2.384 & 0.035 &     0.33 &   1.26 \\
  $-$11.12 &  $-$8.064 & 0.017 &   1.725 & 0.016 &     0.38 &   1.69 \\
  $-$11.14 &  $-$7.507 & 0.013 &   2.123 & 0.018 &     1.06 &   1.90 & 19 \\
  $-$11.15 &    $-$11.970 & 0.020 &  $-$3.590 & 0.030 &     0.19 &   0.21 \\
  $-$12.00 &  $-$4.770 & 0.020 &   2.730 & 0.050 &     0.06 &   0.21 \\
  $-$12.10 &  $-$5.800 & 0.028 &   1.890 & 0.041 &     0.07 &   0.42 \\
  $-$12.18 &  $-$5.281 & 0.013 &   2.356 & 0.061 &     0.14 &   1.05 & 16 \\
  $-$13.56 &    $-$13.565 & 0.014 &   1.097 & 0.075 &     0.12 &   1.05 & 13 \\
  $-$13.82 &    $-$14.674 & 0.015 &   0.558 & 0.026 &     0.14 &   0.84 \\
  $-$13.84 &    $-$14.096 & 0.012 &   0.918 & 0.030 &     0.32 &   1.27 & 12 \\
  $-$14.73 &  10.950 & 0.020 &   8.360 & 0.050 &     0.07 &   0.21 \\
  $-$14.79 &   9.823 & 0.028 &   7.621 & 0.042 &     0.09 &   0.63 \\
  $-$14.85 &  10.383 & 0.016 &   8.055 & 0.015 &     0.18 &   0.84 & 11 \\
  $-$15.33 &    $-$14.683 & 0.043 &   0.806 & 0.053 &     0.09 &   0.63 & 9 \\
  $-$15.88 &    $-$16.248 & 0.025 &   0.386 & 0.015 &     0.24 &   1.05 \\
  $-$16.88 &    $-$16.111 & 0.049 &   0.402 & 0.022 &     0.24 &   1.26 \\
  $-$16.01 &    $-$16.783 & 0.075 &  $-$0.011 & 0.056 &     0.10 &   1.05 & 8 \\
  $-$16.00 &    $-$15.730 & 0.021 &   0.640 & 0.030 &     0.10 &   0.21 \\
  $-$16.84 &    $-$15.630 & 0.020 &   0.720 & 0.040 &     0.08 &   0.21 \\
  $-$16.93 &    $-$16.671 & 0.038 &  $-$0.044 & 0.041 &     0.10 &   0.84 \\
  $-$17.88 &    $-$15.399 & 0.028 &   0.698 & 0.086 &     0.07 &   1.05 & 5 \\
  $-$17.99 &    $-$16.494 & 0.023 &   0.015 & 0.027 &     0.10 &   0.84 \\
  $-$18.11 &    $-$15.923 & 0.028 &   0.427 & 0.014 &     0.21 &   1.26 \\
  $-$19.11 &    $-$14.146 & 0.033 &   0.719 & 0.054 &     0.08 &   0.63 & 4 \\
  $-$22.07 &   3.157 & 0.058 &  11.527 & 0.093 &     0.17 &   1.47 & 3 \\
  $-$22.08 &   2.585 & 0.023 &  11.114 & 0.038 &     0.09 &   1.05 \\
 $-$118.02 &   $-$133.740 & 0.020 &  16.390 & 0.061 &     0.07 &   1.05 & 2 \\
 $-$120.52 &   $-$134.040 & 0.030 &  16.820 & 0.060 &     0.06 &   0.21 & 1 \\
 \hline  
\multicolumn{8}{c}{03/10/25} \\ \hline
  $-$10.56 &    $-$10.531 & 0.099 &  $-$1.231 & 0.037 &     0.35 &   0.84 & 20 \\
  $-$11.24 &    $-$10.240 & 0.017 &  $-$1.126 & 0.016 &     0.60 &   0.63 & 18 \\
  $-$11.42 &  $-$8.595 & 0.015 &  $-$0.683 & 0.010 &     1.01 &   1.47 & 19 \\
  $-$11.65 &  $-$5.900 & 0.015 &  $-$0.248 & 0.025 &     0.39 &   1.05 & 16 \\
  $-$12.31 &    $-$12.052 & 0.028 &  $-$1.641 & 0.050 &     0.11 &   0.63 \\
  $-$12.33 &   6.045 & 0.025 &   7.452 & 0.031 &     0.14 &   0.63 & 15 \\
  $-$12.97 &    $-$14.908 & 0.018 &  $-$1.729 & 0.024 &     0.26 &   1.05 & 13 \\
  $-$14.74 &   9.660 & 0.020 &   5.221 & 0.041 &     0.15 &   0.42 & 11 \\
  $-$14.83 &    $-$16.260 & 0.010 &  $-$2.660 & 0.040 &     0.17 &   0.21 \\
  $-$15.27 &    $-$15.732 & 0.082 &  $-$1.958 & 0.030 &     0.86 &   2.32 & 9 \\
  $-$15.46 &    $-$12.570 & 0.031 &  $-$5.380 & 0.040 &     0.14 &   0.21 \\
  $-$18.08 &    $-$16.476 & 0.051 &  $-$2.213 & 0.043 &     0.09 &   0.84 & 5 \\
  $-$117.46 &   $-$137.101 & 0.010 &  13.176 & 0.052 &     0.13 &   0.42 \\
  $-$117.49 &   $-$136.557 & 0.001 &  13.857 & 0.018 &     0.53 &   1.05 & 2 \\
  $-$117.56 &   $-$136.010 & 0.020 &  14.420 & 0.050 &     0.13 &   0.21 \\
  $-$119.12 &   $-$136.947 & 0.020 &  14.304 & 0.044 &     0.12 &   0.84 & 1 \\
 \hline
\multicolumn{8}{c}{03/12/20} \\ \hline
  $-$11.18 &    $-$10.449 & 0.022 &  $-$1.612 & 0.047 &     0.46 &   0.84 & 18 \\
  $-$11.25 &  $-$9.970 & 0.030 &  $-$1.700 & 0.040 &     0.20 &   0.21 & 17 \\
  $-$11.56 &  $-$6.014 & 0.025 &  $-$0.733 & 0.064 &     0.22 &   0.42 & 16 \\
  $-$11.57 &  $-$8.839 & 0.029 &  $-$0.903 & 0.102 &     0.21 &   0.84 & 19 \\
  $-$12.39 &   5.898 & 0.025 &   6.980 & 0.047 &     0.18 &   0.42 & 15 \\
  $-$11.98 &    $-$10.353 & 0.037 &  $-$1.573 & 0.025 &     0.38 &   1.05 \\
  $-$12.81 &    $-$15.144 & 0.034 &  $-$2.205 & 0.029 &     0.27 &   0.84 & 13 \\
  $-$12.84 &    $-$14.686 & 0.032 &  $-$2.407 & 0.071 &     0.14 &   0.42 & 14 \\
  $-$13.01 &    $-$10.233 & 0.018 &  $-$1.571 & 0.037 &     0.24 &   0.84 \\
  $-$14.41 &    $-$20.610 & 0.030 &    $-$17.410 & 0.040 &     0.11 &   0.21 \\
  $-$14.88 &    $-$15.720 & 0.045 &  $-$2.447 & 0.038 &     0.79 &   1.69 & 9 \\
  $-$15.80 &    $-$16.175 & 0.023 &  $-$2.800 & 0.053 &     0.15 &   1.47 & 7 \\
  $-$16.15 &    $-$16.696 & 0.034 &  $-$2.639 & 0.080 &     0.20 &   1.48 & 6 \\
 $-$117.40 &   $-$137.195 & 0.010 &  13.362 & 0.052 &     0.16 &   0.84 & 2 \\
 $-$119.43 &   $-$137.537 & 0.010 &  13.832 & 0.068 &     0.13 &   1.47 & 1 \\
 \hline
\multicolumn{8}{c}{04/02/09} \\ \hline
  $-$10.41 &    $-$10.530 & 0.096 &  $-$2.376 & 0.071 &     0.11 &   0.63 & 20 \\
  $-$11.19 &    $-$10.414 & 0.035 &  $-$2.392 & 0.029 &     0.23 &   1.05 & 18 \\
  $-$11.26 &  $-$5.996 & 0.008 &  $-$1.470 & 0.026 &     0.33 &   1.05 & 16 \\
  $-$11.26 &  $-$6.470 & 0.023 &  $-$2.066 & 0.074 &     0.14 &   0.84 \\
  $-$11.32 &  $-$5.522 & 0.059 &  $-$1.178 & 0.035 &     0.18 &   1.05 \\
  $-$11.35 &  $-$9.917 & 0.023 &  $-$2.133 & 0.033 &     0.23 &   0.84 & 17 \\
  $-$11.46 &    $-$10.950 & 0.020 &  $-$2.840 & 0.040 &     0.15 &   0.21 \\
  $-$12.26 &   5.954 & 0.027 &   6.127 & 0.074 &     0.11 &   0.63 & 15 \\
  $-$12.31 &    $-$14.460 & 0.030 &  $-$2.710 & 0.050 &     0.10 &   0.21 \\
  $-$12.39 &   6.417 & 0.028 &   6.437 & 0.043 &     0.12 &   0.42 \\
  $-$12.70 &    $-$14.692 & 0.037 &  $-$2.791 & 0.053 &     0.17 &   1.26 & 14 \\
  $-$12.95 &    $-$15.181 & 0.017 &  $-$3.003 & 0.032 &     0.23 &   1.05 & 13 \\
  $-$12.95 &    $-$15.666 & 0.032 &  $-$3.514 & 0.062 &     0.10 &   0.84 \\
  $-$14.41 &    $-$16.630 & 0.020 &  $-$3.940 & 0.050 &     0.12 &   0.21 \\
  $-$14.76 &    $-$16.172 & 0.022 &  $-$3.358 & 0.031 &     0.42 &   1.26 & 9 \\
  $-$14.85 &    $-$15.177 & 0.017 &  $-$2.802 & 0.034 &     0.24 &   1.05 \\
  $-$14.96 &    $-$15.669 & 0.025 &  $-$3.051 & 0.074 &     0.38 &   2.11 & 10 \\
  $-$15.83 &    $-$16.660 & 0.015 &  $-$3.469 & 0.018 &     0.49 &   1.68 & 6 \\
  $-$15.91 &    $-$17.165 & 0.019 &  $-$3.991 & 0.036 &     0.19 &   1.26 \\
  $-$16.01 &    $-$16.159 & 0.039 &  $-$3.224 & 0.026 &     0.30 &   1.47 & 7 \\
 $-$119.55 &   $-$137.920 & 0.099 &  13.133 & 0.063 &     0.11 &   1.26 & 1 \\
 $-$119.64 &   $-$137.512 & 0.010 &  13.374 & 0.089 &     0.11 &   0.84 \\
 \hline 
\multicolumn{8}{c}{04/04/26} \\ \hline
  $-$10.20 &    $-$10.844 & 0.007 &  $-$3.223 & 0.026 &     0.61 &   1.27 & 20 \\
  $-$11.09 &  $-$6.161 & 0.017 &  $-$2.245 & 0.015 &     0.29 &   1.26 & 16 \\
  $-$12.24 &    $-$15.368 & 0.030 &  $-$3.753 & 0.012 &     1.08 &   1.26 & 13 \\
  $-$12.31 &   5.730 & 0.030 &   5.290 & 0.040 &     0.26 &   0.21 & 15 \\
  $-$12.31 &    $-$14.950 & 0.010 &  $-$3.270 & 0.040 &     0.27 &   0.21 \\
  $-$12.56 &    $-$10.454 & 0.014 &  $-$3.167 & 0.046 &     0.28 &   1.68 & 18 \\
  $-$12.88 &    $-$14.984 & 0.045 &  $-$3.742 & 0.031 &     0.27 &   0.84 & 14 \\
  $-$14.25 &    $-$15.846 & 0.031 &  $-$3.846 & 0.022 &     0.47 &   0.84 &10 \\
  $-$15.04 &    $-$16.178 & 0.053 &  $-$3.768 & 0.026 &     0.34 &   1.26 & 9 \\
  $-$15.22 &    $-$16.704 & 0.058 &  $-$4.160 & 0.009 &     1.14 &   2.53 & 6 \\
 $-$117.83 &   $-$128.972 & 0.010 &  12.126 & 0.022 &     0.26 &   1.26 \\
 $-$119.62 &   $-$138.876 & 0.010 &  12.325 & 0.041 &     0.27 &   1.68 & 1 \\
\enddata

\tablenotetext{1}{Local-standard-of-rest velocity at the intensity peak in 
units of km~s$^{-1}$.}
\tablenotetext{2}{Position offset with respect to the delay-tracking center 
at data correlation ($\alpha_{J2000}=$19$^{h}$15$^{m}$35$^{s}$\hspace{-2pt}.2150,
$\delta_{J2000}=$~$+$21$^{\circ}$36$^{\prime}$33\arcsec.900) in units of mas.}
\tablenotetext{3}{Peak intensity of the feature in units of Jy~beam$^{-1}$.}
\tablenotetext{4}{Full velocity width of maser emission in units of \kms. 
The minimum is equal to the velocity spacing of a spectral channel (0.21\kms).}
\tablenotetext{5}{Identification of the maser feature listed in table 3. The feature is designated as \iras1913:I2007 {\it N}, where {\it N} is the ordinal 
source number given in this column (I2007 stands for sources found by 
Imai et al. and listed in 2007). }
\end{deluxetable}

\end{document}